\documentclass[prd]{revtex4}
\usepackage{color}

\begin{document}
\title{Mass Correction to Chiral Kinetic Equations}
\author{Ziyue Wang$^1$, Xingyu Guo$^2$, Shuzhe Shi$^3$ and Pengfei Zhuang$^1$}
\affiliation{$^1$Physics Department, Tsinghua University, Beijing 100084, China\\
             $^2$Institute of Quantum Matter, South China Normal University, Guangzhou 510006, China\\
             $^3$Department of Physics, McGill University, 3600 University Street, Montreal, QC, H3A 2T8, Canada}
\date{\today}
\begin{abstract}
We study fermion mass correction to chiral kinetic equations in electromagnetic fields. Different from the chiral limit where fermion number density is the only independent distribution, the number and spin densities are coupled to each other for massive fermion systems. To the first order in $\hbar$, we derived the quantum correction to the classical on-shell condition and the Boltzmann-type transport equations. To the linear order in the fermion mass, the mass correction does not change the structure of the chiral kinetic equations and behaves like additional collision terms. While the mass correction exists already at classical level in general electromagnetic fields, it is only a first order quantum correction in the study of chiral magnetic effect.
\end{abstract}
\maketitle

\section{Introduction}
The chiral anomaly of Quantum Chromodynamics (QCD) or Quantum Electrodynamics (QED) is recently widely discussed both theoretically and experimentally. Putting a system of chiral fermions in an external magnetic field, the chiral imbalance between the left and right handed fermions leads to an electric current along the direction of the magnetic field. It is called Chiral Magnetic Effect~\cite{son_quantum_2004,metlitski_anomalous_2005,kharzeev_parity_2006,kharzeev_charge_2007,fukushima_chiral_2008} and triggered a lot of interests in nuclear physics~\cite{kharzeev_effects_2008,liao_anomalous_2015,kharzeev_chiral_2016} and condensed matter physics~\cite{fedorov_chiral_2016}. Three ingredients are crucial for the generation of the chiral magnetic effect, the magnetic field, the presence of chiral imbalance, and the massless fermions. In high energy heavy ion collisions which are expected to be a way of realizing the chiral magnetic effect, the coexistence of the first two ingredients may occur in the quark matter created in the initial stage of the collisions. However, all quarks in QCD are massive, even in extremely hot quark matter. To check the degree of chiral anomaly in a real fermion system, it is necessary to study the fermion mass effect on the chiral magnetic effect. This is not a trivial problem even in the case of small fermion mass. With nonzero mass, fermions with different helicity are coupled to each other, and the fermion field contains four components instead of two components for Weyl fermions in chiral limit. It is of fundamental necessity to find out how finite mass modifies the chiral anomaly effects. There are already several attempts to investigate the mass effect on chiral imbalance~\cite{copinger_axial_2018,hou,lin,hattori} and non-Abelian Berry curvature~\cite{gao_relativistic_2019,weickgenannt_kinetic_2019}.

In high energy nuclear collisions, the possible chiral magnetic effect should carry highly non-equilibrium nature, indicated by the magnetized quark matter in non-equilibrium state created in the very beginning of the collisions. For an out-of-equilibrium system, a natural way to describe the transport phenomena is through the kinetic theory in Wigner function formalism~\cite{deGroot,elze}. The chiral magnetic effect in out-of-equilibrium state in chiral limit is recently widely studied in the framework of kinetic theory~\cite{hidaka_nonlinear_2018,hidaka_non-equilibrium_2018-1,hidaka_non-equilibrium_2018}. By applying semiclassical expansion method to the kinetic equations, to the first order in $\hbar$, the chiral anomaly related effects are incorporated into the transport equation for the chiral fermion distribution function~\cite{gao_covariant_2017,huang_complete_2018,gao_disentangling_2018}. The transport equation is also applied to phenomenologically study the charge separation in the pre-thermal stage of heavy ion collisions~\cite{huang_out--equilibrium_2018}.

In this paper, we generally study the fermion mass correction to the chiral kinetic equations in external electromagnetic fields in equal-time Wigner function formalism. In Section \ref{s2}, we first review the kinetic equations for the spin components of the equal-time Wigner function and their semiclassical expansion in $\hbar$, and then focus on the quantum correction to the classical on-shell condition and the Boltzmann equations to the first order in $\hbar$. In Section \ref{s3}, we derive the transport equations for the chiral components and take Taylor expansion in quark mass to explicitly see the mass correction to the chiral kinetic equations. In Section \ref{s4}, we take the mass correction to the CME flow as an example of the general kinetic theory and analytically solve the transport equation. We summarize the study in Section \ref{s5}.

\section{Equal-time kinetic equations}
\label {s2}
The moving of charged fermions in external electromagnetic fields is controlled by the QED Lagrangian density
\begin{equation}
\label{lagrangian}
\mathcal{L}=\bar{\psi}(i\gamma^\mu D_\mu-m)\psi-\frac{1}{4}F^{\mu\nu}F_{\mu\nu},
\end{equation}
where $m$ is the fermion mass, $F_{\mu\nu}=\partial_\mu A_\nu-\partial_\nu A_\mu$ is the electromagnetic field tensor, and the covariant derivative $D_\mu=\partial_\mu+iQA_\mu$ couples the quark field $\psi$ with electric charge $Q$ to the electromagnetic fields $A_\mu$. The covariant fermion Wigner function $W(x,p)$ is defined as the ensemble average of the Wigner operator in vacuum state, and the Wigner operator is the four-dimensional Fourier transform of the covariant density matrix~\cite{vasak_quantum_1987},
\begin{equation}
\label{w4}
W(x,p)=\int d^4y e^{ipy}\langle\psi(x_+)e^{iQ\int_{-1/2}^{1/2}dsA(x+sy)y}\bar\psi(x_-)\rangle,
\end{equation}
where the exponential function of the electromagnetic fields is the gauge link between the two points $x_\pm =x\pm y/2$ which guarantees the gauge invariance of the kinetic theory. It is easy to see that, the Wigner function is the analogy to the probability distribution in quantum mechanics. When the gauge fields are external fields, the gauge link can be taken out from the ensemble average $\langle ... \rangle$.

To extract particle distribution functions from the Wigner function and solve the kinetic equations as an initial value problem, one usually introduce the equal-time Wigner function~\cite{bialynicki-birula_phase-space_1991}
\begin{equation}
\label{w3}
{\cal W}(x,{\bf p})=\int d^3y e^{ipy}\langle\psi(x_+)e^{iQ\int_{-1/2}^{1/2}dsA(x+sy)y}\psi^\dag(x_-)\rangle
\end{equation}
with $y=(0,{\bf y})$. It is clear that, the equal-time Wigner function is not Lorentz covariant and the two Wigner functions are related to each other through the energy integration,
\begin{equation}
\label{w34}
{\cal W}(x,{\bf p}) =\int dp_0 W(x,p)\gamma^0.
\end{equation}
The two Wigner functions are equivalent to each other only when the particles are on the energy shell. In quantum off-shell case, the covariant Wigner function is equivalent to the collection of all the energy moments $\int dp_0 p_0^n W(x,p)\gamma^0$ with $n=0,1,2,...$. The equal-time Wigner function ${\cal W}(x,{\bf p})$ is only the zeroth order energy moment of the covariant one $W(x,p)$. We will see soon the hierarchy among all the energy moments in general quantum case.

The covariant and equal-time kinetic equations in external electromagnetic fields are systematically investigated by Vasak, Gyulassy, and Elze~\cite{vasak_quantum_1987}, Bialynicki-Birula, Gornicki, and Rafelski~\cite{bialynicki-birula_phase-space_1991} and Zhuang and Heinz~\cite{zhuang_relativistic_1996}. Since neither the covariant nor the equal-time Wigner functions are real, the physical phase-space densities are defined through their spin components,
\begin{eqnarray}
\label{decom}
W &=& \frac{1}{4}\left[F+i\gamma_5 P+\gamma_\mu V^\mu+\gamma_\mu\gamma_5 A^\mu+\frac{1}{2}\sigma_{\mu\nu}S^{\mu\nu}\right],\nonumber\\
{\cal W} &=& \frac{1}{4}\left[f_0+\gamma_5 f_1-i\gamma_0\gamma_5 f_2+\gamma_0 f_3+\gamma_5\gamma_0{\bf \gamma}\cdot{\bf g}_0+\gamma_0{\bf \gamma}\cdot{\bf g}_1-i{\bf \gamma}\cdot {\bf g}_2-\gamma_5{\bf \gamma}\cdot {\bf g}_3\right].
\end{eqnarray}
By calculating the physical densities of the system like charge, energy, momentum and angular momentum in terms of the equal-time Wigner function, one can establish the physical meaning of the equal-time components~\cite{bialynicki-birula_phase-space_1991}. For instance, $f_0$ is the charge density, $f_3$ the mass density, ${\bf g}_0$ the spin current, ${\bf g}_1$ the number current, and ${\bf g}_3$ the intrinsic magnetic moment. Taking the derivatives of the density matrix $\psi(x_+)\bar\psi(x_-)$ with respect to $x$ and $y$ and using the Dirac equations controlling the motion of the quark fields $\psi$ and $\bar\psi$, one derives the kinetic equations for the $16$ covariant spin components~\cite{vasak_quantum_1987},
\begin{eqnarray}
\label{kinetic4}
&& \Pi^\mu V_\mu=mF,\nonumber\\
&& \hbar D^\mu A_\mu = 2mP,\nonumber\\
&& 2\Pi_\mu F-\hbar D^\nu S_{\nu\mu} = 2mV_\mu,\nonumber\\
&& -\hbar D_\mu P+\epsilon_{\mu\nu\sigma\rho}\Pi^\nu S^{\sigma\rho}=2m A_\mu,\nonumber\\
&& \hbar(D_\mu V_\nu- D_\nu V_\mu)+2\epsilon_{\mu\nu\sigma\rho}\Pi^\sigma A^\rho=2mS_{\mu\nu},\nonumber\\
&& \hbar D^\mu V_\mu = 0,\nonumber\\
&& \Pi^\mu A_\mu=0,\nonumber\\
&& \hbar D_\mu F = -2\Pi^\nu S_{\nu\mu},\nonumber\\
&& 4\Pi^\mu P=-\hbar\epsilon_{\mu\nu\sigma\rho}D^\nu S^{\sigma\rho},\nonumber\\
&& 2(\Pi_\mu V_\nu-\Pi_\nu V)=\hbar\epsilon_{\mu\nu\sigma\rho} D^\sigma A^\rho,
\end{eqnarray}
where the covariant derivative $D_\mu$ and generalized momentum $\Pi_\mu$ in phase space are defined as
\begin{eqnarray}
\label{dmupimu4}
D_\mu(x,p)&=& \partial_\mu-Q\int_{-1/2}^{1/2}ds F_{\mu\nu}(x-i\hbar s\partial_p)\partial_p^{\nu},\nonumber\\
\Pi_\mu(x,p) &=& p_\mu-iQ\hbar\int_{-1/2}^{1/2}ds s F_{\mu\nu}(x-i\hbar s\partial_p)\partial_p^{\nu}.
\end{eqnarray}
We have explicitly shown the $\hbar$-dependence here in order to be able to discuss the semiclassical expansion of the kinetic equations in the following. We now take the relation (\ref{w34}) between ${\cal W}(x,{\bf p})$ and $W(x,p)$. By doing $p_0-$integration of the covariant equations (\ref{kinetic4}), one obtains the equal-time transport equations which are the extension of the classical Boltzmann equation~\cite{zhuang_relativistic_1996,guo_out--equilibrium_2018},
\begin{eqnarray}
\label{transport}
&&\hbar(D_t f_0+{\bf D}\cdot{\bf g}_1) = 0,\nonumber\\
&&\hbar(D_t f_1+{\bf D}\cdot{\bf g}_0) = -2m f_2,\nonumber\\
&&\hbar D_t f_2-2{\bf \Pi}\cdot{\bf g}_3 =2m f_1,\nonumber\\
&&\hbar D_t f_3-2{\bf \Pi}\cdot{\bf g}_2 = 0,\nonumber\\
&&\hbar(D_t{\bf g}_0+{\bf D} f_1)-2{\bf \Pi}\times{\bf g}_1 = 0,\nonumber\\
&&\hbar(D_t{\bf g}_1+{\bf D} f_0)-2{\bf \Pi}\times{\bf g}_0 = -2m{\bf g}_2,\nonumber\\
&&\hbar(D_t{\bf g}_2-{\bf D}\times{\bf g}_3)+2{\bf \Pi} f_3 =2m{\bf g}_1,\nonumber\\
&&\hbar(D_t{\bf g}_3+{\bf D}\times{\bf g}_2)+2{\bf \Pi} f_2 = 0,
\end{eqnarray}
and the equal-time constraint equations which are the extension of the classical on-shell condition~\cite{zhuang_relativistic_1996,guo_out--equilibrium_2018},
\begin{eqnarray}
\label{constraint}
&&\int dp_0 p_0 V_0-{\bf \Pi}\cdot{\bf g}_1+\Pi_0 f_0 = mf_3,\nonumber\\
&&\int dp_0 p_0 A_0+{\bf \Pi}\cdot{\bf g}_0-\Pi_0f_1 = 0,\nonumber\\
&&\int dp_0 p_0 P+\frac{1}{2}\hbar{\bf D}\cdot{\bf g}_3+\Pi_0f_2 = 0,\nonumber\\
&&\int dp_0 p_0 F-\frac{1}{2}\hbar{\bf D}\cdot{\bf g}_2+\Pi_0f_3 = mf_0,\nonumber\\
&&\int dp_0 p_0 {\bf A}+\frac{1}{2}\hbar{\bf D}\times{\bf g}_1+{\bf \Pi}f_1-\Pi_0{\bf g}_0 = -m{\bf g}_3,\nonumber\\
&&\int dp_0 p_0 {\bf V}-\frac{1}{2}\hbar{\bf D}\times{\bf g}_0+{\bf \Pi}f_0-\Pi_0{\bf g}_1 = 0,\nonumber\\
&&\int dp_0 p_0 S^{0i}{\bf e}_i-\frac{1}{2}\hbar{\bf D}f_3+{\bf \Pi}\times{\bf g}_3-\Pi_0{\bf g}_2 = 0,\nonumber\\
&&\int dp_0 p_0 S_{jk}\epsilon^{jki}{\bf e}_i-\hbar {\bf D}f_2+2{\bf \Pi}\times{\bf g}_2+2\Pi_0{\bf g}_3 = 2m{\bf g}_0,
\end{eqnarray}
where the equal-time operators are defined as
\begin{eqnarray}
\label{dmupimu3}
D_t &=& \partial_t+Q\int_{-1/2}^{1/2}ds {\bf E}({\bf x}+is\hbar{\bf \nabla}_p)\cdot{\bf \nabla}_p,\nonumber\\
{\bf D} &=& {\bf \nabla}+Q\int_{-1/2}^{1/2}ds{\bf B}({\bf x}+is\hbar{\bf \nabla}_p)\times{\bf \nabla}_p,\nonumber\\
\Pi_0 &=& iQ\hbar\int_{-1/2}^{1/2}dss{\bf E}({\bf x}+is\hbar{\bf \nabla}_p)\cdot{\bf \nabla}_p,\nonumber\\
{\bf \Pi} &=& {\bf p} -iQ\hbar\int_{-1/2}^{1/2}dss{\bf B}({\bf x}+is\hbar{\bf \nabla}_p)\times{\bf \nabla}_p.
\end{eqnarray}
We have here directly used the electromagnetic field strengths ${\bf E}$ and ${\bf B}$ instead of the fields $A_\mu$. It is clear that, the constraint equations couple the equal-time components $f_i(x,{\bf p})$ and ${\bf g}_i(x,{\bf p})\ (i=0,1,2,3)$ with the first order energy moments $\int dp_0 p_0 \Gamma_a(x,p)\gamma_0$ with $\Gamma_a=\{F, P, V_\mu, A_\mu, S_{\mu\nu}\}$. Only in the classical limit with on-shell condition $p_0=\pm E_p$, the first order moments are reduced to $\pm E_p \{f_i,{\bf g}_i\}$, and the transport and constraint equations become a group of closed kinetic equations for the equal-time Wigner function. In general case with quantum off-shell effect, all the energy moments are independent, they couple to each other and form a hierarchy of kinetic equations~\cite{zhuang_equal-time_1998,ochs_wigner_1998}.

To see explicitly the classical limit and quantum correction order by order, we now make semiclassical expansions for the covariant and equal-time components and operators,
\begin{eqnarray}
\label{hbarexpansion}
&& F = F^{(0)}+\hbar F^{(1)}+\cdots,\nonumber\\
&& P = P^{(0)}+\hbar P^{(1)}+\cdots,\nonumber\\
&& V_\mu = V_\mu^{(0)}+\hbar V_\mu^{(1)}+\cdots,\nonumber\\
&& A_\mu = A_\mu^{(0)}+\hbar A_\mu^{(1)}+\cdots,\nonumber\\
&& S_{\mu\nu} = S_{\mu\nu}^{(0)}+\hbar S_{\mu\nu}^{(1)}+\cdots,\nonumber\\
&& f_i = f_i^{(0)}+\hbar f_i^{(1)}+\cdots,\nonumber\\
&& {\bf g}_i = {\bf g}_i^{(0)}+\hbar {\bf g}_i^{(1)}+\cdots,\nonumber\\
&& D_t = D_t^{(0)}+\hbar D_t^{(1)}+\cdots,\nonumber\\
&& {\bf D} = {\bf D}^{(0)}+\hbar{\bf D}^{(1)}+\cdots,\nonumber\\
&& \Pi_0 = \Pi_0^{(0)}+\hbar\Pi_0^{(1)}+\cdots,\nonumber\\
&& {\bf \Pi} = {\bf \Pi}^{(0)}+\hbar{\bf \Pi}^{(1)}+\cdots,\nonumber\\
&& D_t^{(0)}=\partial_t+Q{\bf E}\cdot{\bf \nabla}_p,\nonumber\\
&& {\bf D}^{(0)}={\bf \nabla}+Q{\bf B}\times{\bf \nabla}_p,\nonumber\\
&& \Pi_0^{(0)}=0,\nonumber\\
&& {\bf \Pi}^{(0)}={\bf p}.
\end{eqnarray}

Substituting the expansions into the equal-time transport and constraint equations, we first consider the classical limit with $\hbar = 0$. Taking the classical on-shell condition for the positive and negative energy parts of the Wigner function $W(x,p)=W^+(x,p)\delta(p_0-E_p)+W^-(x,p)\delta(p_0+E_p)$, the constraint equations (\ref{constraint}) automatically determine the position of the shell, namely the particle energy $E_p=\sqrt{m^2+{\bf p}^2}$ and reduce the number of independent spin components. In general quantum case, all the $16$ spin components are independent. In the classical limit, however, only the fermion number density $f_0^{(0)}$ and spin current ${\bf g}_0^{(0)}$ are independent, and the other components can simply be expressed in terms of them~\cite{zhuang_relativistic_1996,guo_out--equilibrium_2018},
\begin{eqnarray}
\label{relation0}
f_1^{(0)\pm} &=& \pm{{\bf p}\over E_p}\cdot{\bf g}^{(0)\pm}_0,\nonumber\\
f_2^{(0)\pm} &=& 0,\nonumber\\
f_3^{(0)\pm} &=& \pm{m\over E_p}f^{(0)\pm}_0,\nonumber\\
{\bf g}_1^{(0)\pm} &=& \pm {{\bf p}\over E_p}f^{(0)\pm}_0,\nonumber\\
{\bf g}_2^{(0)\pm} &=& {{\bf p}\times{\bf g}^{(0)\pm}_0\over m},\nonumber\\
{\bf g}_3^{(0)\pm} &=& \mp{E_p^2 {\bf g}^{(0)\pm}_0-({\bf p}\cdot{\bf g}^{(0)\pm}_0){\bf p}\over mE_p}.
\end{eqnarray}
Note that, the classical limit of the transport equations (\ref{transport}) provides only a part of the above relations and does not contribute any new information.

To include quantum correction to the first order in $\hbar$, a straightforward idea is the extension of the on-shell condition, $W(x,p)=W^{+}(x,p)\delta(p_0-E_p-\hbar\delta E_p)+W^{-}(x,p)\delta(p_0+E_p+\hbar\delta E_p)$. The particles are still on the shell, but the position of the shell is shifted from $E_p$ to $E_p+\hbar \delta E_p$, where $\delta E_p$ is a spin-independent shell shift induced by the quantum correction. Using the $\hbar$-expansion for the $\delta$ function $\delta(p_0-E_p-\hbar\delta E_p)=\delta(p_0-E_p)-\hbar\delta E_p\delta'(p_0-E_p)$ and doing the integrations $\int dp_0 p_0 \Gamma_a^{(0)\pm}(x,p)\delta'(p_0\mp E_p)$ by parts, the constraint equations (\ref{constraint}) at the first order in $\hbar$ become
\begin{eqnarray}
\label{hbar1}
&& \pm E_p f_0^{(1)\pm}+\Delta E_{p0}^\pm -{\bf p}\cdot {\bf g}_1^{(1)}=mf_3^{(1)},\nonumber\\
&& \pm E_p f_1^{(1)\pm}+\Delta E_{p1}^\pm -{\bf p}\cdot {\bf g}_0^{(1)}=0,\nonumber\\
&& \pm E_p f_2^{(1)\pm}+\Delta E_{p2}^\pm -{1\over 2}{\bf D}^{(0)}\cdot {\bf g}_3^{(0)}=0,\nonumber\\
&& \pm E_p f_3^{(1)\pm}+\Delta E_{p3}^\pm  -{1\over 2}{\bf D}^{(0)}\cdot {\bf g}_2^{(0)}=mf_0^{(1)},\nonumber\\
&& \pm E_p {\bf g}_0^{(1)\pm}+\Delta{\bf E}_{p0}^\pm -{\bf p}f_1^{(1)}-{1\over 2}{\bf D}^{(0)}\times {\bf g}_1^{(0)}=m{\bf g}_3^{(1)},\nonumber\\
&& \pm E_p {\bf g}_1^{(1)\pm}+\Delta{\bf E}_{p1}^\pm -{\bf p}f_0^{(1)}-{1\over 2}{\bf D}^{(0)}\times {\bf g}_0^{(0)}=0,\nonumber\\
&& \pm E_p {\bf g}_2^{(1)\pm}+\Delta{\bf E}_{p2}^\pm +{\bf p}\times {\bf g}_3^{(1)}+{1\over 2}{\bf D}^{(0)}f_3^{(0)}=0,\nonumber\\
&& \pm E_p {\bf g}_3^{(1)\pm}+\Delta{\bf E}_{p3}^\pm -{\bf p}\times {\bf g}_2^{(1)}=m{\bf g}_0^{(1)}
\end{eqnarray}
with the definition $\Delta E_{pi}^\pm(x,{\bf p})=\delta E_p f_i^{(0)\pm}(x,{\bf p})$ and $\Delta {\bf E}_{pi}^\pm(x,{\bf p})=\delta E_p {\bf g}_i^{(0)\pm}(x,{\bf p})$ controlled by the shell shift and classical components. Since $f_i^{(0)}$ and ${\bf g}_i^{(0)}$ should satisfy the classical constraints (\ref{relation0}), it is impossible to find a shift $\delta E_p$ which satisfies the $16$ first-order constraints (\ref{hbar1}). This means that, when quantum correction is included, there is no more an energy shell for the particles. We also tried component dependent shell shifts $\delta E_{pa}$ by assuming $\Gamma_a(x,p)=\Gamma_a^{+}(x,p)\delta(p_0-E_p-\hbar\delta E_{pa})+\Gamma_a^{-}(x,p)\delta(p_0+E_p+\hbar\delta E_{pa})$. In this case, the constraint equations (\ref{hbar1}) are still valid, but $\delta E_p$ in $\Delta E_{pi}^\pm$ and $\Delta {\bf E}_{pi}^\pm$ is replaced by $\delta E_{pa}$. Again, we cannot work out $\Delta E_{pi}^\pm$ and $\Delta {\bf E}_{pi}^\pm$ which satisfy both the classical and first-order constraints (\ref{relation0}) and (\ref{hbar1}).

The spin component dependent shell at $E_p+\delta E_{pa}$ is not a real energy shell for particles, it is already a specific expression of the off-shell effect. However, it does not fulfil the kinetic equations. To include a general off-shell effect in the kinetic theory, we add a continuous function of $p_0$ to the classical on-shell condition, namely we take
\begin{equation}
\label{offshell}
\Gamma_a(x,p) = \Gamma_a^+(x,p)\left(\delta(p_0-E_p)-\hbar{\cal A}(p)\right)+\Gamma_a^-(x,p)\left(\delta(p_0+E_p)+\hbar{\cal A}(p)\right),
\end{equation}
where the spectral function ${\cal A}(p)$ is a quantum correction to classical particles. By substituting the covariant components $\Gamma_a$ into the original constraint equations (\ref{constraint}), we obtain again their first order equations (\ref{hbar1}) with $\Delta E_{pa}$ characterized by the continuous spectrum ${\cal A}(p)$, $\Delta E_{pa} = \int dp_0 p_0\Gamma_a^{(0)}{\cal A}(p)$. Note that, the $\Gamma_a^{(0)}$ multiplied by the spectral function ${\cal A}(p)$ are not on the shell. Therefore, $\Delta E_{pa}$ are not constrained by the classical relations (\ref{relation0}), they are controlled only by the constraint equations (\ref{hbar1}). By eliminating the first-order components in (\ref{hbar1}), we obtain $\Delta E_{pa}$ in the local rest frame,
\begin{eqnarray}
\Delta E_{p0}^\pm &=& \mp{{\bf B}\cdot{\bf g}^{(0)\pm}_0\over 2E_p},\nonumber\\
\Delta E_{p1}^\pm &=& -{{\bf B}\cdot{\bf p}\over 2E^2_p}f_0^{(0)\pm},\nonumber\\
\Delta E_{p2}^\pm &=& {{\bf E}\cdot{\bf g}^{(0)\pm}_0\over 2m},\nonumber\\
\Delta E_{p3}^\pm &=& \pm{{\bf p}\cdot({\bf E}\times{\bf g}^{(0)\pm}_0)\over 2mE_p}-{{\bf B}\cdot{\bf g}^{(0)\pm}_0\over 2m}+{({\bf B}\cdot{\bf p})({\bf p}\cdot{\bf g}^{(0)\pm}_0)\over 2mE^2_p},\nonumber\\
\Delta{\bf E}_{p0}^\pm &=& \pm\left(\mp{{\bf E}\times{\bf p}\over 2E^2_p}+m{{\bf B}\over 2E_p}\right)f_0^{(0)\pm},\nonumber\\
\Delta{\bf E}_{p1}^\pm &=& \mp{{\bf E}\times{\bf g}^{(0)\pm}_0\over 2E_p}-{{\bf B}({\bf p}\cdot{\bf g}^{(0)\pm}_0)\over 2E^2_p},\nonumber\\
\Delta{\bf E}_{p2}^\pm &=& {m{\bf E}\over 2E^2_p}f_0^{(0)\pm},\nonumber\\
\Delta{\bf E}_{p3}^\pm &=& {m{\bf B}\over 2E^2_p}f_0^{(0)\pm}.
\end{eqnarray}

The constraint equations (\ref{hbar1}) not only determine the quantum correction to the classical mass shell, but also reduce the number of independent spin components at the first order in $\hbar$. Similar to the classical case, $f_0^{(1)}$ and ${\bf g}_0^{(1)}$ are still the independent spin components, and the other components are determined by them and their classical limit,
\begin{eqnarray}
\label{relation1}
f^{(1)\pm}_1 &=& \pm\frac{{\bf p}\cdot{\bf g}^{(1)\pm}_0}{E_p}\pm\frac{{\bf p}\cdot{\bf B}}{2E^3_p}f^{(0)\pm}_0,\nonumber\\
f^{(1)\pm}_2 &=&-{{\bf D}^{(0)}\cdot{\bf g}^{(0)\pm}_0\over 2m}+{{\bf p}\cdot({\bf p}\cdot{\bf D}^{(0)}){\bf g}^{(0)\pm}_0\over 2mE_p^2}-{({\bf B}\times{\bf p})\cdot{\bf g}^{(0)\pm}_0\over mE_p^2}\mp{{\bf E}\cdot{\bf g}^{(0)\pm}_0\over 2mE_p}, \nonumber\\
f^{(1)\pm}_3 &=& \pm\frac{mf_0^{(1)}}{E_p}\mp{({\bf p}\times {\bf D}^{(0)})\cdot{\bf g}^{(0)\pm}_0\over 2mE_p}+{{\bf p}\cdot({\bf E}\times{\bf g}^{(0)\pm}_0)\over 2mE^2_p}\mp{{\bf B}\cdot{\bf g}^{(0)\pm}_0\over 2mE_p}\mp{({\bf B}\cdot{\bf p})({\bf p}\cdot{\bf g}^{(0)\pm}_0)\over 2mE^3_p} , \nonumber\\
{\bf g}^{(1)\pm}_1 &=& \pm\frac{{\bf p}}{E_p} f^{(1)}_0\pm\frac{1}{2E_p}{\bf D}^{(0)}\times{\bf g}^{(0)}_0+\frac{{\bf E}}{2E^2_p}\times{\bf g}^{(0)\pm}_0\pm\frac{{\bf B}({\bf p}\cdot{\bf g}^{(0)\pm}_0)}{2E^3_p},\nonumber\\
{\bf g}^{(1)\pm}_2 &=& \frac{{\bf p}\times{\bf g}_0^{(1)\pm}}{m}\pm\left(\frac{{\bf p}({\bf p}\cdot{\bf E})}{2mE_p^3}-\frac{{\bf E}}{2mE_p}\right)f_0^{(0)\pm}+\frac{{\bf p}}{2mE_p^2}{\bf p}\cdot{\bf D}^{(0)} f_0^{(0)\pm}-\frac{1}{2m}{\bf D}^{(0)} f_0^{(0)\pm},\nonumber\\
{\bf g}^{(1)\pm}_3 &=& \mp\left(\frac{E_p}{m}{\bf g}_0^{(1)\pm}-\frac{{\bf p}\cdot {\bf g}_0^{(1)\pm}}{m E_p}{\bf p}\right)+\left(\frac{{\bf E}\times{\bf p}}{2mE_p^2}\mp\frac{m {\bf B}}{2E^3_p}\right)f^{(0)\pm}_0\mp\frac{1}{2m E_p}{\bf p}\times{\bf D}^{(0)} f^{(0)\pm}_0.
\end{eqnarray}

We now calculate the dynamical equations controlling the evolution of the independent spin components $f_0$ and ${\bf g}_0$ at classical level and including the first order quantum correction. At classical level, the behavior of $f_0^{(0)}$ and ${\bf g}_0^{(0)}$ is controlled by the transport equations (\ref{transport}) to the first order in $\hbar$,
\begin{eqnarray}
\label{habar1-2}
&& D_t^{(0)} f_0^{(0)}+{\bf D}^{(0)}\cdot{\bf g}_1^{(0)}=0,\nonumber\\
&& D_t^{(0)} f_1^{(0)}+{\bf D}^{(0)}\cdot{\bf g}_0^{(0)}=-2mf_2^{(1)},\nonumber\\
&& {\bf p}\cdot{\bf g}_3^{(1)}=-mf_1^{(1)},\nonumber\\
&& D_t^{(0)} f_3^{(0)}-2{\bf p}\cdot{\bf g}_2^{(1)}=0,\nonumber\\
&& D_t^{(0)} {\bf g}_0^{(0)}+{\bf D}^{(0)}f_1^{(0)}-2{\bf p}\times{\bf g}_1^{(1)}=0,\nonumber\\
&& D_t^{(0)} {\bf g}_1^{(0)}+{\bf D}^{(0)}f_0^{(0)}-2{\bf p}\times{\bf g}_0^{(1)}=-2m{\bf g}_2^{(1)},\nonumber\\
&& D_t^{(0)} {\bf g}_2^{(0)}-{\bf D}^{(0)}\times {\bf g}_3^{(0)}+2{\bf p}f_3^{(1)}=2m{\bf g}_1^{(1)},\nonumber\\
&& D_t^{(0)} {\bf g}_3^{(0)}-{\bf D}^{(0)}\times {\bf g}_2^{(0)}+2{\bf p}f_2^{(1)}=0.
\end{eqnarray}
Substituting the classical relation between ${\bf g}_1^{(0)}$ and $f_0^{(0)}$ into the first equation, it leads to the Boltzmann-type transport equation for the particle number density $f_0^{(0)}$,
\begin{equation}
\label{transportf0}
\left(D_t^{(0)} \pm {{\bf p}\over E_p}\cdot{\bf D}^{(0)}\right)f^{(0)\pm}_0 = 0.
\end{equation}
Combining the second and the last equations to eliminate the first-order component $f_2^{(1)}$, and then taking into account the classical relations between $f_1^{(0)}, {\bf g}_2^{(0)}, {\bf g}_3^{(0)}$ and ${\bf g}_0^{(0)}$, we obtain the second Boltzmann-type transport equation for the particle spin density ${\bf g}_0^{(0)}$,
\begin{equation}
\label{transportg0}
\left(D_t^{(0)} \pm {{\bf p}\over E_p}\cdot{\bf D}^{(0)}\right){\bf g}^{(0)\pm}_0 = {1 \over E_p^2}\left[{\bf p}\times \left({\bf E}\times {\bf g}^{(0)\pm}_0\right)\mp E_p{\bf B}\times{\bf g}^{(0)\pm}_0\right]
\end{equation}
which is the phase-space version of a generalized Bargmann-Michel-Telegdi equation~\cite{bargmann_precession_1959,zhuang_relativistic_1996-1} and describes spin precession in external electromagnetic fields. It is clear that, the particle number density $f_0^{(0)}$ and spin density ${\bf g}_0^{(0)}$ are independent to each other, they are not coupled in the transport equations. Since we do not include interaction among particles in the Lagrangian density, there is no collision term on the right-hand side of the transport equation for $f_0^{(0)}$. However, for the spin density ${\bf g}_0^{(0)}$, the interaction between spin angular momentum and electromagnetic fields results in collision terms in the transport equation.

The dynamical evolution of the particle number density $f_0^{(1)}$ and spin density ${\bf g}_0^{(1)}$ to the first order in $\hbar$ is controlled by the transport equations (\ref{transport}) to the second order in $\hbar$. Taking the classical and first-order constraints (\ref{relation0}) and (\ref{relation1}) and using the classical transport equations (\ref{transportf0}) and (\ref{transportg0}), a straightforward but tedious calculation leads to
\begin{eqnarray}
\label{transport1}
\left(D_t^{(0)}\pm\frac{{\bf p}}{E_p} \cdot {\bf D}^{(0)}\right) f^{(1)\pm}_0 &=& \frac{{\bf E}}{2E_p^2}\cdot{\bf D}^{(0)}\times{\bf g}_0^{(0)\pm}\mp\frac{1}{2E_p^3}{\bf B}\cdot({\bf p}\cdot{\bf D}^{(0)}){\bf g}_0^{(0)\pm}+\frac{{\bf B}\times{\bf p}}{E_p^4}\cdot {\bf E}\times{\bf g}_0^{(0)\pm},\nonumber\\
\left(D_t^{(0)}\pm\frac{{\bf p}}{E_p}\cdot{\bf D}^{(0)}\right){\bf g}^{(1)\pm}_0 &=& {1 \over E_p^2}\left[{\bf p}\times \left({\bf E}\times {\bf g}^{(1)\pm}_0\right)\mp E_p{\bf B}\times{\bf g}^{(1)\pm}_0\right]\mp \left(\frac{\bf B}{2E_p^3}\pm\frac{{\bf E}\times{\bf p}}{2E_p^4}\right){\bf p}\cdot{\bf D}^{(0)} f^{(0)\pm}_0\nonumber\\
&&\mp \left(\frac{({\bf p}\cdot{\bf E})({\bf E}\times{\bf p})}{E^5_p}\pm\frac{{\bf p}\times({\bf B}\times{\bf E})}{2E_p^4}\right)f^{(0)\pm}_0.
\end{eqnarray}
It is obvious that, the number density $f_0$ which comes from the covariant vector component $V_\mu$ and the spin density ${\bf g}_0$ which comes from the covariant axial vector component $A_\mu$ are coupled to each other at quantum level. There are now collisions terms in the transport equation for the number density $f_0^{(1)}$ due to the spin interaction with the electromagnetic fields. With an appropriate initial condition, one can solve firstly the classical transport equations and then the quantum transport equations order by order. The higher order quantum corrections can be derived in a similar way.

It is not necessary to choose the number density $f_0$ and spin density ${\bf g}_0$ as the independent spin components, this can be seen from the classical and first-order constraints (\ref{relation0}) and (\ref{relation1}). In some time it becomes better to take $f_0$ and the magnetic moment ${\bf g}_3$ as the independent ones, see the next section. In this case, we need the transport equations for ${\bf g}_3^{(0)}$ and ${\bf g}_3^{(1)}$,
\begin{eqnarray}
\label{transportg3}
{\bf p}\cdot\left(D_t^{(0)}\pm {{\bf p}\over E_p}\cdot{\bf D}^{(0)}\right){\bf g}^{(0)\pm}_3 &=& -{{\bf p}^2\over E_p^2}\left({\bf E}\pm {{\bf p}\over E_p}\times{\bf B}\right)\cdot {\bf g}^{(0)\pm}_3\mp {m^2\over E_p^3}{\bf p}\cdot({\bf B}\times {\bf g}^{(0)\pm}_3),\nonumber\\
{\bf p}\cdot\left(D_t^{(0)}\pm {{\bf p}\over E_p}\cdot{\bf D}^{(0)}\right){\bf g}^{(1)\pm}_3 &=& -{{\bf p}^2\over E_p^2}\left({\bf E}\pm {{\bf p}\over E_p}\times{\bf B}\right)\cdot {\bf g}^{(1)\pm}_3\mp {m^2\over E_p^3}{\bf p}\cdot({\bf B}\times {\bf g}^{(1)\pm}_3)\nonumber\\
&&+{m\over 2E_p^4}({\bf p}\cdot{\bf B})({\bf p}\cdot{\bf D}^{(0)}) f_0^{(0)\pm}\mp{m\over 2E_p^3}{\bf p}\cdot({\bf E}\times{\bf D}^{(0)}) f^{(0)\pm}_0\nonumber\\
&&\pm {3m\over 2E_p^5}({\bf p}\cdot {\bf B})({\bf p}\cdot{\bf E})f^{(0)\pm}_{0}\pm{m{\bf p}^2\over 2E_p^5}({\bf B}\cdot{\bf E})f^{(0)\pm}_0.
\end{eqnarray}

\section {Transport equations for chiral components}
\label {s3}
In chiral limit, while the vector and axial vector currents $V_\mu$ and $A_\mu$ are coupled to each other, their combinations $J_\mu=V_\mu+A_\mu$ and $J_\mu=V_\mu-A_\mu$ are decoupled. The physics behind is the number conservation of left-handed and right-handed fermions. To see the mass correction to the chiral conservation, we still introduce the chiral currents $J_\mu^\chi=V_\mu+\chi A_\mu\ (\chi=\pm)$ in covariant formalism or $f_\chi=f_0+\chi f_1$ and ${\bf g}_\chi={\bf g}_1+\chi{\bf g}_0$ in equal-time formalism. In chiral limit, $J_\mu^\chi$ represent the currents of fermions with definite chirality. From the classical and quantum relations (\ref{relation0}) and (\ref{relation1}), the zeroth and first order chiral components ${\bf g}_\chi$ can be expressed as
\begin{eqnarray}
\label{relation2}
{\bf g}_\chi^{(0)\pm} &=& {\bf g}_1^{(0)\pm}+\chi{\bf g}_0^{(0)\pm}\nonumber\\
                      &=& \pm{{\bf p}\over E_p}f^{(0)\pm}_\chi\mp \chi\frac{m}{E_p}{\bf g}^{(0)\pm}_3,\nonumber\\
{\bf g}_\chi^{(1)\pm} &=& {\bf g}_1^{(1)\pm}+\chi{\bf g}_0^{(1)\pm}\nonumber\\
                      &=& \pm{{\bf p}\over E_p}f^{(1)\pm}_\chi-\frac{\chi}{2}\left(\frac{{\bf p}({\bf p}\cdot{\bf B})}{E^4_p}+\frac{m^2{\bf B}}{E^4_p}\pm\frac{{\bf p}\times{\bf
                          E}}{E^3_p}\pm{{\bf p}\cdot{\bf B}\over E_p^3}+\frac{{\bf p}}{E_p^2}\times{\bf D}^{(0)}\right)f_\chi^{(0)\pm}\nonumber\\
                      &&\mp\chi\frac{m}{E_p}{\bf g}^{(1)\pm}_3\mp\frac{m{\bf E}\times{\bf g}^{(0)\pm}_3}{2E^3_p}-\frac{m{\bf D}^{(0)}\times{\bf g}^{(0)\pm}_3}{2E^2_p}-\frac{m}{2E_p^4}{\bf g}^{(0)\pm}_3\times({\bf B}\times{\bf p}).
\end{eqnarray}
In chiral limit with $m=0$, ${\bf g}_\chi^{(0)}=\frac{{\bf p}}{E_p}f_\chi^{(0)}$ and ${\bf g}_\chi^{(1)}$ is a linear combination of $f_\chi^{(0)}$ and $f_\chi^{(1)}$. The two degrees of freedom for a massive fermion system (number $f_\chi$ and current ${\bf g}_\chi$) are reduced to one for a massless fermion system $(f_\chi)$. For massless fermions with certain chirality, the spin is not an independent degree of freedom but always parallel or anti-parallel to the momentum, and the spin distribution can be determined by the number density. For massive fermions, ${\bf g}_\chi$ and $f_\chi$ are, however, independent components, as the spin direction does not follow the momentum direction. In this case, ${\bf g}_\chi$ is related to not only $f_\chi$ but also the magnetic moment ${\bf g}_3$ or the spin density ${\bf g}_0$.

Using the transport equations for $f_0^{(0)\pm}$ and $f_0^{(1)\pm}$ derived in Section \ref{s2}, we have
\begin{eqnarray}
{\bf p}\left(D_t^{(0)} \pm {{\bf p}\over E_p}\cdot{\bf D}^{(0)}\right)f_1^{(0)\pm} &=& m\left[\left(D_t^{(0)} \pm {{\bf p}\over E_p}\cdot{\bf D}^{(0)}\right){\bf g}^{(0)\pm}_3\pm\frac{1}{E_p}{\bf B}\times{\bf g}^{(0)\pm}_3\right],\nonumber\\
{\bf p}\left(D_t^{(0)}\pm\frac{\bf p}{E_p}\cdot{\bf D}^{(0)}\right)f_1^{(1)\pm} &=& m\left[\left(D_t^{(0)}\pm\frac{\bf p}{E_p}\cdot{\bf D}^{(0)}\right){\bf g}_3^{(1)\pm}\pm\frac{1}{E_p}{\bf B}\times{\bf g}_3^{(1)\pm}\right]\nonumber\\
&&\pm\left[\frac{\bf p}{2E_p^3}{\bf p}\cdot({\bf E}\times{\bf D}^{(0)})+\frac{m^2}{2E_p^3}{\bf E}\times{\bf D}^{(0)}\mp\left(\frac{{\bf p}({\bf p}\cdot{\bf B})}{2E_p^4}+\frac{m^2{\bf B}}{2E_p^4}\right){\bf p}\cdot{\bf D}^{(0)}\right] f_0^{(0)\pm}\nonumber\\
&&\pm\left(\frac{({\bf E}\cdot{\bf B}){\bf p}}{2E_p^3}-\frac{3({\bf p}\cdot{\bf E})({\bf p}\cdot{\bf B}){\bf p}}{2E_p^5}-\frac{3m^2({\bf p}\cdot{\bf E}){\bf B}}{2E_p^5}\right)f_0^{(0)\pm}
\end{eqnarray}
for $f_1^{(0)\pm}$ and $f_1^{(1)\pm}$ and then the transport equations for the classical and quantum chiral components $f_\chi^{(0)\pm}$ and $\tilde f_\chi^{(1)\pm}=f_\chi^{(1)\pm}\mp\chi{{\bf p}\cdot{\bf B}\over 2E_p^3}f_\chi^{(0)\pm}$,
\begin{eqnarray}
\label{transport3}
&&\left(D_t^{(0)}\pm\frac{{\bf p}}{E_p}\cdot{\bf D}^{(0)}\right)f_\chi^{(0)\pm} = \chi mF_1[{\bf g}_3^{(0)}],\\
&&\left(D_t^{(0)}\pm\frac{\bf p}{E_p}\cdot{\bf D}^{(0)}\right)\tilde f_\chi^{(1)\pm}\pm\chi\frac{{\bf p}\cdot{\bf B}}{2E_p^3}D_t^{(0)}f_\chi^{(0)\pm}+\chi\left(\frac{{\bf p}({\bf p}\cdot{\bf B})}{E_p^4}\pm\frac{{\bf E}\times{\bf p}}{2E_p^3}\right)\cdot{\bf D}^{(0)} f_\chi^{(0)\pm}=\chi mF_1[{\bf g}_3^{(1)}]+mF_2[{\bf g}_3^{(0)}],\nonumber
\end{eqnarray}
where we have shifted the first-order distribution from $f_\chi^{(1)}$ to $\tilde f_\chi^{(1)}$ to remove the infrared divergence in chiral limit~\cite{hidaka_nonlinear_2018,chen}, and the two functions of the magnetic moment ${\bf g}_3$ are defined as
\begin{eqnarray}
\label{f12}
F_1[{\bf g}_3] &=& -\frac{{\bf E}\cdot{\bf g}_3^\pm}{p^2},\nonumber\\
F_2[{\bf g}_3] &=& \pm\frac{1}{2p^3}{\bf D}^{(0)}\cdot({\bf E}\times{\bf g}_3^\pm)+\frac{1}{2p^4}({\bf p}\cdot{\bf D}^{(0)})({\bf B}\cdot{\bf g}_3^\pm)\mp\frac{3}{2p^5}({\bf p}\times{\bf B})\cdot({\bf E}\times{\bf g}_3^\pm).
\end{eqnarray}
To see clearly the mass correction to the chiral kinetic equations, we have taken Taylor expansion in terms of the fermion mass $m$ in the transport equations (\ref{transport3}) and kept only the linear terms in $m$ which are explicitly shown on the right-hand side.

The sum of the two equations in (\ref{transport3}) leads to the transport equation for the chiral component $f_\chi=f_\chi^{(0)}+\hbar\tilde f_\chi^{(1)}$,
\begin{equation}
\left(D_t^{(0)}\pm\frac{\bf p}{p}\cdot{\bf D}^{(0)}\right)f_\chi^{\pm}\pm\chi\hbar\frac{{\bf p}\cdot{\bf B}}{2p^3}D_t^{(0)}f_\chi^{\pm}+\chi\hbar\left(\frac{{\bf p}({\bf p}\cdot{\bf B})}{p^4}\pm\frac{{\bf E}\times{\bf p}}{2p^3}\right)\cdot{\bf D}^{(0)} f_\chi^{\pm} = \chi mF_1[{\bf g}_3^{\pm}]+\hbar mF_2[{\bf g}_3^{(0)\pm}].
\end{equation}
We should emphasize again that, different from the chiral limit, the transport equations for the magnetic moment ${\bf g}_3$ listed in the end of the last section are needed to close this kinetic equation for the chiral component $f_\chi$.

Taking homogeneous electromagnetic fields ${\bf E}$ and ${\bf B}$ to simplify the equations and to compare with the known results in chiral limit and introducing berry curvature~\cite{berry_quantal_1984} ${\bf b}=\chi{\bf p}/2p^3$, dispersion relation $\epsilon_p=p(1-\hbar Q{\bf B}\cdot{\bf b})$ and velocity ${\bf v}_p={\bf \nabla}_p\epsilon_p={\bf p}/p(1+2\hbar Q{\bf b}\cdot{\bf B})-\hbar Q({\bf p}/p\cdot{\bf b}){\bf B}$, the transport equation can be simplified as
\begin{equation}
\label{cktm}
\partial_tf_\chi^\pm+\dot{\bf x}\cdot{\bf \nabla} f_\chi^\pm+\dot{\bf p}\cdot{\bf \nabla}_p f_\chi^\pm = \chi m{F_1[{\bf g}_3^\pm]\over \sqrt G}+\hbar m\frac{F_2[{\bf g}_3^{(0)\pm}]}{\sqrt{G}}
\end{equation}
with the phase-space factor $G=(1+\hbar Q{\bf B}\cdot{\bf b})^2$ and the equations of motion
\begin{eqnarray}
\label{motion}
\dot{\bf x} &=& \frac{1}{\sqrt G}\left[{\bf v}_p+\hbar Q\left({\bf v}_p\cdot{\bf b}\right){\bf B}+\hbar Q{\bf E}\times{\bf b}\right],\nonumber\\
\dot{\bf p} &=& \frac{Q}{\sqrt G}\left[{\bf v}_p\times{\bf B}+{\bf E}+\hbar\left({\bf E}\cdot{\bf B}\right){\bf b}\right].
\end{eqnarray}

In comparison with the chiral kinetic equation for massless fermions~\cite{son_berry_2012,son_kinetic_2013},
\begin{equation}
\label{ckt}
\partial_tf_\chi^\pm+\dot{\bf x}\cdot{\bf \nabla} f_\chi^\pm+\dot{\bf p}\cdot{\bf \nabla}_p f_\chi^\pm = 0,
\end{equation}
the two kinetic equations with and without fermion mass have the same structure: the berry curvature, the equations of motion, and the phase-space factor are exactly the same. The only difference is the nonzero collision terms on the right-hand side generated by the interaction between the massive particle spin and electromagnetic fields.

\section{Solution to the mass correction}
\label {s4}
Given the above kinetic equation (\ref{cktm}) for fermion systems with small mass, it is of great interest to find possible analytic solutions. The great complexity in general case, as we pointed out above, is the mass induced coupling between the chiral component $f_\chi$ and the other independent distribution ${\bf g}_3$. However, when we turn off the electric field ${\bf E}$ and keep only the magnetic field ${\bf B}$, corresponding to the physics of chiral magnetic effect, the effective collision term with $F_1$ which is coupled to ${\bf g}_3$ vanishes, and the other collision term is only related to the classical distribution ${\bf g}_3^{(0)}$ which can be solved through the classical transport equation before. In this case, the collision term $\hbar mF_2[{\bf g}_3^{(0)}]/\sqrt G$ in the linear non-homogenous differential equation is known and the equation can be analytically solved~\cite{yan}.

Considering the three independent vectors ${\bf B}$, ${\bf p}$ and ${\bf B}\times {\bf p}$ in the case with only magnetic field, ${\bf g}_3^{(0)}$ should include three components parallel to the three elementary vectors. From the explicit expression of $F_2$, see (\ref{f12}), the last component does not contribute to the collision term, and we can assume a general form ${\bf g}^{(0)}_3=G_1{\bf p}+G_2{\bf B}$ with two scalar functions $G_1$ and $G_2$. It is also easy to see the disappearance of the momentum derivative in the collision term, $({\bf p}\cdot({\bf B}\times{\bf \nabla}_p))({\bf B}\cdot{\bf g}_3^{(0)})=0$. Finally, the transport equation can be simplified as
\begin{equation}
\label{ktm}
\partial_tf_\chi^\pm+\dot{\bf x}\cdot{\bf \nabla} f_\chi^\pm+\dot{\bf p}\cdot{\bf \nabla}_p f_\chi^\pm = \hbar m\frac{1}{2\sqrt G p^4}({\bf p}\cdot{\bf \nabla})({\bf B}\cdot{\bf g}_3^{(0)\pm}).
\end{equation}

The other point in the case with only magnetic field is that, the mass correction is only a quantum correction, since the collision term is at the first order in $\hbar$. This leads to the conclusion that, the mass correction to the chiral magnetic effect should be small. When the electric field is turned on, the mass correction appears already at classical level, see the first collision term with $F_1$ in (\ref{cktm}). Therefore, in the case with only electrical field or both electrical and magnetic fields, the mass correction will become more important.

We first consider the collisionless limit, namely the chiral kinetic equation (\ref{ckt}). In this limit, the particles will simply undergo free-streaming according to the trajectory determined from the equations of motion (\ref{motion}). Note that such trajectory is different from the usual classical trajectory due to the anomalous terms. For a particle with initial position ${\bf x}_0$ and initial momentum ${\bf p}_0$ at time $t_0$, its position ${\bf x}({\bf x}_0,{\bf p}_0,t_0;t)$ and momentum ${\bf p}({\bf x}_0,{\bf p}_0,t_0;t)$ at time $t$ are given by~\cite{huang_out--equilibrium_2018}
\begin{eqnarray}
x &=& x_0+{1\over QB}\left[p_{x0}\sin\theta+p_{y0}(1-\cos\theta)\right],\nonumber\\
y &=& y_0+{1\over QB}\left[-p_{x0}(1-\cos\theta)+p_{y0}\sin\theta\right],\nonumber\\
z &=& z_0+{\zeta\over \sqrt G p}p_{z0}(t-t_0),\nonumber\\
p_x &=& p_{x0}\cos\theta+p_{y0}\sin\theta,\nonumber\\
p_y &=& -p_{x0}\sin\theta+p_{y0}\cos\theta,\nonumber\\
p_z &=& p_{z0}
\end{eqnarray}
with the definition $\zeta=1+\chi{QBp_z\over p^3}$ and $\theta={\zeta QB\over \sqrt G p}(t-t_0)$, where we have assumed a space and time independent magnetic field along $z$-axis ${\bf B}=B{\bf e}_z$.

Equivalently, a particle found to have position ${\bf x}$ and momentum ${\bf p}$ at time $t$ can be traced back to a state of ${\bf x}_0({\bf x},{\bf p},t;t_0)$ and ${\bf p}_0({\bf x},{\bf p},t;t_0)$ at initial time $t_0$,
\begin{eqnarray}
x_0 &=& x-{1\over QB}\left[p_x\sin\theta-p_y(1-\cos\theta)\right],\nonumber\\
y_0 &=& y-{1\over QB}\left[p_x(1-\cos\theta)+p_y\sin\theta\right],\nonumber\\
z_0 &=& z-{\zeta\over \sqrt G p}p_z(t-t_0),\nonumber\\
p_{x0} &=& p_x\cos\theta-p_y\sin\theta,\nonumber\\
p_{y0} &=& p_x\sin\theta+p_y\cos\theta,\nonumber\\
p_{z0} &=& p_z.
\end{eqnarray}
Therefore, given an initial condition $f_{\chi 0}({\bf x}_0,{\bf p}_0,t_0)$, the solution of the chiral kinetic equation is simply
\begin{equation}
f_\chi^\pm({\bf x},{\bf p},t)=f_{\chi 0}^\pm({\bf x}_0({\bf x},{\bf p},t;t_0),{\bf p}_0({\bf x},{\bf p},t;t_0),t_0).
\end{equation}

We now solve the non-homogeneous differential equation, namely the kinetic equation (\ref{ktm}) with known mass-induced collision term $\beta({\bf x},{\bf p},t)\equiv {\hbar m\over 2\sqrt G p^4}({\bf p}\cdot{\bf \nabla})({\bf B}\cdot{\bf g}_3^{(0)\pm})$. The solution can be analytically written as
\begin{equation}
f_\chi^\pm({\bf x},{\bf p},t) = f_{\chi 0}^\pm({\bf x}_0({\bf x},{\bf p},t;t_0),{\bf p}_0({\bf x},{\bf p},t;t_0),t_0)+\int_{t_0}^t\beta({\bf x}({\bf x}_0,{\bf p}_0,t_0;t'),{\bf p}({\bf x}_0,{\bf p}_0,t_0,t'),t')dt'.
\end{equation}
The first term is the solution of the corresponding homogeneous differential equation, namely the solution of the chiral kinetic equation in chiral limit, and the second term is the correction from the small fermion mass.

\section{Summary}
\label {s5}
While the quantum chiral anomaly and related phenomena in fermion systems are widely discussed in chiral limit, the mass correction in real case should be seriously considered. For a non-equilibrium system, a systematic way to study quantum correction to particle transport phenomena is the kinetic theory in Wigner function formalism. In this paper we systematically studied the fermion mass correction to the chiral kinetic equations in external electromagnetic fields in the frame of equal-time transport theory.

In chiral limit, fermions are always on mass shell, although quantum correction may change the position of the shell. For massive fermions, the on-shell condition is no longer a solution of the equal-time constraint equations at quantum level, and the off-shell effect should be included in the quantum kinetic theory. At first order in $\hbar$, we fixed the off-shell induced terms by analytically solving the constraint equations. With the help of these constraints, we derived the transport equations for the particle number and spin densities at classical level and to the first order quantum correction. To see clearly the fermion mass correction to the chiral kinetic equations, we take Taylor expansion in terms of the mass, and to the linear order we obtained kinetic equations for massive fermions. The mass correction is reflected as effective collision terms in the transport equations. Different from chiral limit where the chiral number density is the only independent quantity and its transport equation controls the evolution of the system, spin density becomes independent for massive fermions, and the chiral number density and spin density are coupled to each other. Only in the case with only magnetic field, the two densities are decoupled. In this case, the mass correction is a quantum correction, and the chiral number density can be analytically solved. In general case with electric field, the mass correction appears already at classical level, and the effect on chiral properties should be more important.
\\

\appendix {\bf Acknowledgement}: The work is supported by the NSFC grant Nos. 11575093 and 11890712.

\end{document}